\documentclass{article}
\usepackage[dvips]{epsfig}
\begin{document}
\begin{center}
{\large \bf Local chain ordering in amorphous polymer melts: Influence of
  chain stiffness}
\end{center}

\begin{center}
  Roland Faller, Alexander Kolb, and Florian M\"uller-Plathe \\
  \small Max-Planck-Institut f\"ur Polymerforschung, D-55128 Mainz
\end{center}

\hrulefill

  {\bf Abstract} - Molecular dynamics simulation of a generic
  polymer model is applied to study melts of polymers with different
  types of intrinsic stiffness. Important static observables of the
  single chain such as gyration radius or persistence length are
  determined. Additionally we investigate the overall static melt
  structure including pair correlation function, structure function
  and orientational correlation function. 

\hrulefill

\section{Introduction}
The detailed structure of bulk amorphous polymers is a topic of
scientific interest because it is necessary for the microscopic
understanding of their properties. However, because of the amorphous
nature of polymer melts and glasses, structural information is
difficult to obtain  experimentally. 

In particular, it is interesting to know {\it how amorphous} a polymer melt
is on a local scale, i.e. how much residual order is left on a local
scale, how far local order extends before it disappears into the
long-range disorder of amorphous systems \cite{Doi:86.1}, and how the
local order depends on the molecular architecture. The interest has
recently been revived by solid-state NMR studies of Graf et
al. \cite{graf98}, from which it was inferred that a melt of
polybutadiene is far more ordered than hitherto expected. The
alignment of polymer chains is restricted to a local scale, there is
no sign of nematic ordering.  

In order to get a better understanding of local packing and ordering
effects computer simulations are very helpful, because the system is
precisely known and because one has access to all data including
positions and velocities of all particles at all times. Atomistic 
simulations may be useful in order to get an understanding of a
specific system whereas simplified models yield the properties of
generic polymer melts. Additionally, they need much less simulation
time which allows to tackle relatively big systems for long times
\cite{kremer89,puetz98}. Therefore, a simple bead-spring model may be
a good starting point to elaborate generic packing effects.  

There was some work done for semiflexible chains, both by Monte Carlo and
molecular dynamics, mostly to study liquid crystals 
\cite{kolinski86,affouard96,escobedo97,kamien97} or focusing on
confined systems \cite{yethiraj94,hendricks95}. The influence of chain
stiffness on the dynamic structure factors of polymer melts was investigated
also by analytical theory by Harnau {\it et. al} \cite{harnau96} who
found major discrepancies to the fully flexible system for large
scattering vectors, i.e. on short distances. 

In a previous article \cite{faller98b}, we showed that there is
considerable local chain alignment even in melts of fully flexible chains 
(persistence length: 1 monomer diameter). This persistence length originates
from excluded volume interaction. If there was no interaction at all (except
for connectivity) the persistence length would be zero (e.g. polycatenans).
In the present contribution, this model is extended to include some more
information about the chemical architecture of the polymer. We firstly
introduce bending potentials of different strength, in order to study the
effect of semiflexibility of single chain structure as well as on the mutual
local orientation of neighboring chains. Secondly, we study models with
alternating stiffness in an attempt to mimic simplistically polymers with rigid
subunits connected by more flexible links, like polybutadiene and
polyisoprene with their alternating single and double bonds which are
currently under investigation experimentally \cite{graf98}.  

\section{Simulated System}
We performed molecular dynamics simulations (for details of the parallel
program {\it POLY}, see ref. \cite{puetz98}) of melts of polymer chains at a
density $\rho^{*}=0.85$ and temperature $T^{*}=1$ at a timestep $\delta
t^{*}=0.01$ using a truncated and shifted Lennard-Jones potential 
(Weeks-Chandler-Anderson potential) for the excluded-volume interaction
between all beads.  Lennard Jones reduced units are used throughout this
paper where the mass $m$, the potential well depth $\epsilon$ and the
radius of the potential minimum $\sigma$ define the unit system.
\begin{equation}
  V_{LJ}(r)=4\epsilon\left[\left(\frac{\sigma}{r}\right)^{12}-
    \left(\frac{\sigma}{r}\right)^{6}\right]+\epsilon,\quad
  r<r_{cutoff}=\sqrt[6]{2}\sigma 
\end{equation}
and a finitely extendable non-linear elastic (FENE) potential 
\begin{equation}
  V_{FENE}(r)=\frac{\alpha}{2}\frac{R^{2}}{\sigma^{2}}
  \ln\left(1-\frac{r^{2}}{R^{2}} \right),\quad r<R=1.5\sigma,\, \alpha =30
\end{equation}
for the connection of neighboring beads. Additionally, a bond angle potential
\begin{equation}
  V_{angle} =
  x\left(1-\frac{{\bf r}_{i-1,i}\cdot{\bf r}_{i,i+1}}{r_{i-1,i}\,r_{i,i+1}}
  \right)  
\end{equation}
is used. This model system was already widely studied both for flexible
\cite{kremer89,kremer88} and for semiflexible or liquid crystalline polymer
systems \cite{affouard96,hendricks95}. 

To first approximation, there is
$\frac{l_{p}}{l_{b}}=\frac{x}{k_{B}T}$ where $l_{p}$ is the
persistence length (see section 3)and $l_{b}$ the bond length. In our
units, the numerical values for $x$ and $l_{p}$ therefore
coincide. This potential is  applied to every bead, to every 2nd bead,
every 3rd bead etc. The latter is a useful model for polymers with
alternating stiffness such as single-bond, double-bond sequences or
for copolymers with different persistence lengths of the constituents. In the
following, we refer to a system with angular potential strength $x$
and a (topological) distance of $y$ monomers between two applications of the
bond potential as $x$-$y$ system. In this sense, a fully flexible
chain is referred as 0-1 chain. A 5-2 chain for example has $x=5$ applied to
every second bond angle.  

All simulated systems contained 500 chains of 50, 100 or 200 monomers each, so
the overall number of particles was between 25.000 and 100.000 in a cubic
periodic box. 

\begin{figure}
  \includegraphics[width=5cm,angle=-90]{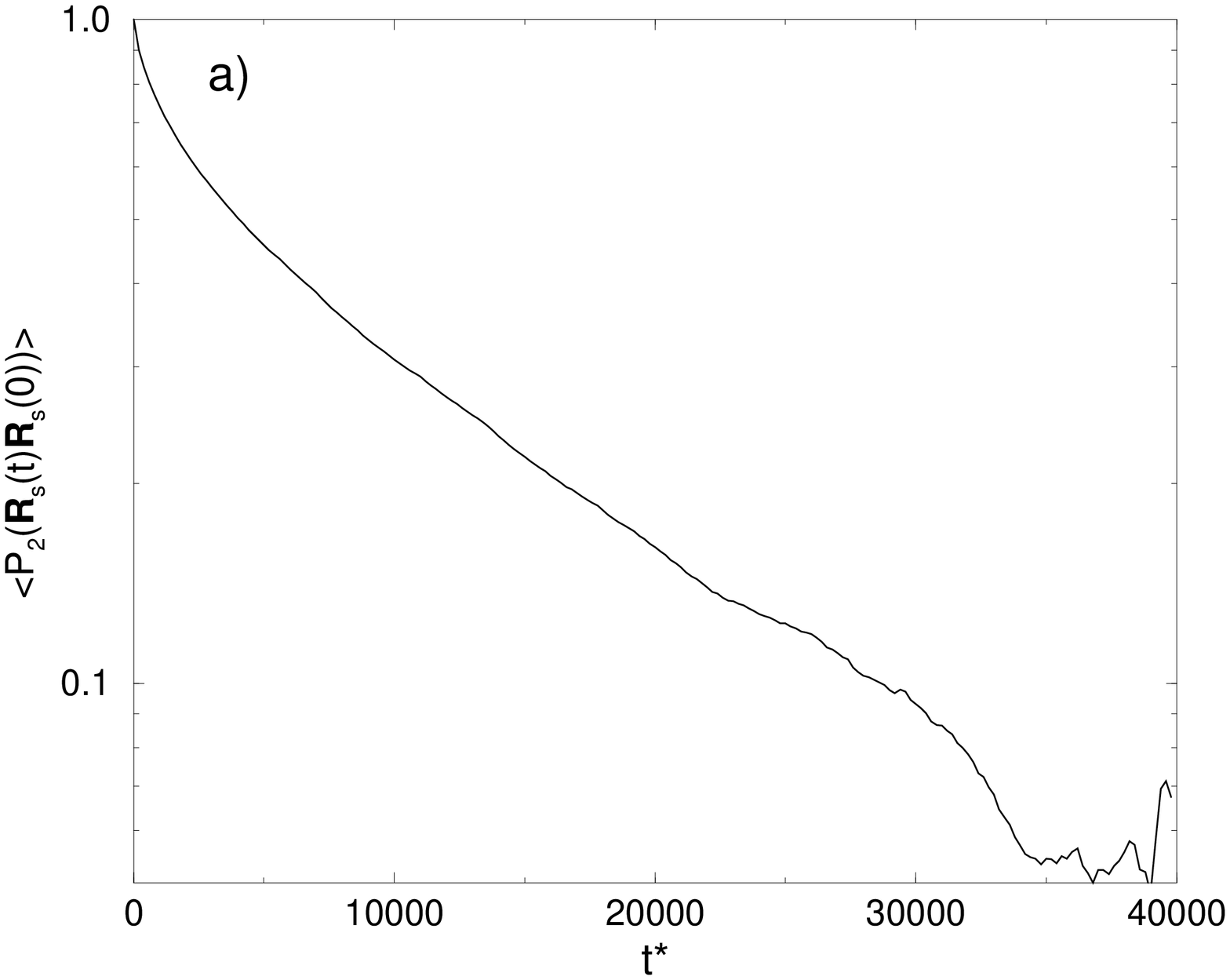}
  \includegraphics[width=5cm,angle=-90]{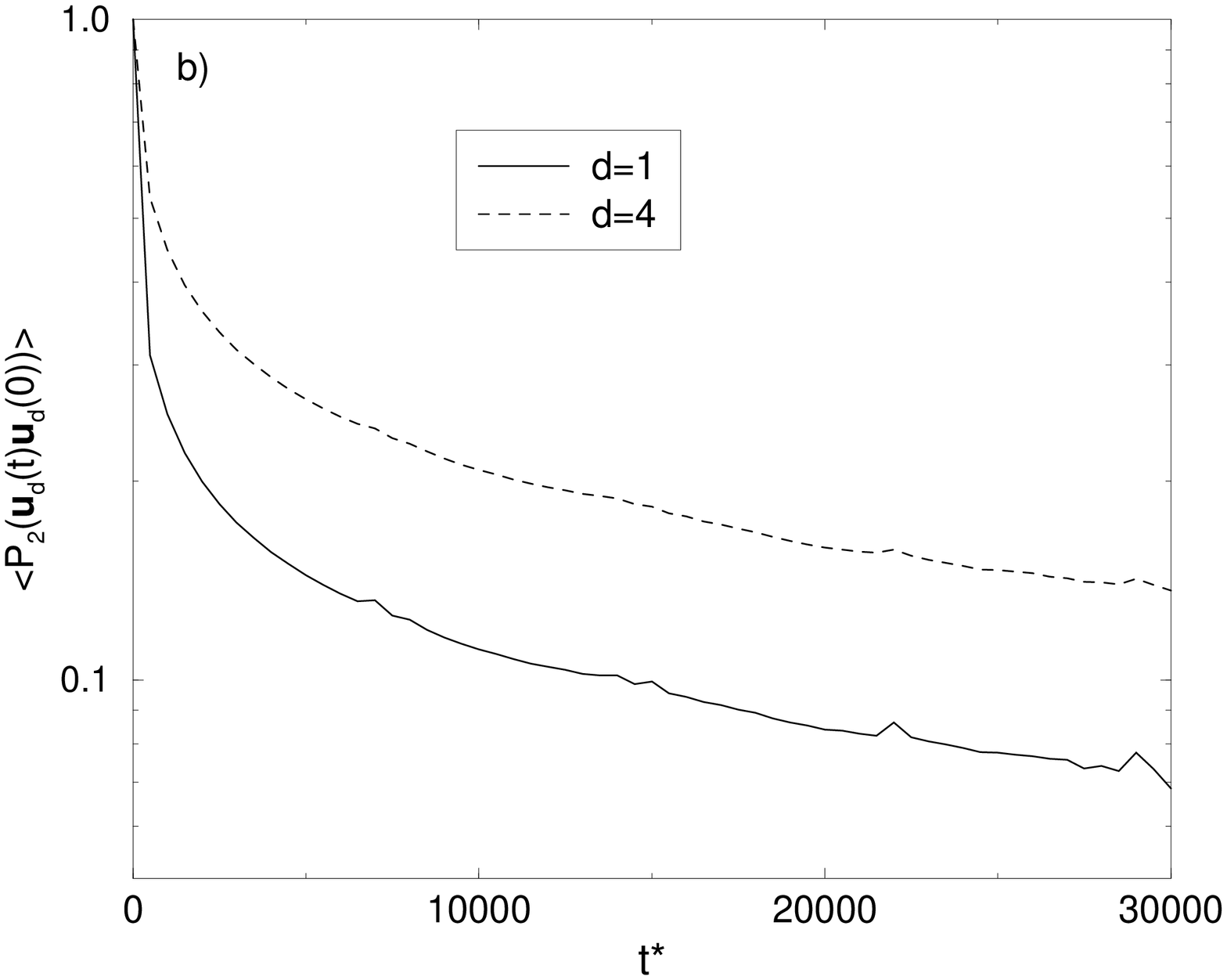}
  \caption{ a) Reorientation of the end-end vector for 5-1 chains (50
    monomers). b) Reorientation of segments of length $d$ ($d+1$ monomers) for
    5-1 chains of length 200. }
  \label{fig:equi}
\end{figure}

The short-chain systems (50 monomers) could be observed until the
auto-correlation function of the end-to-end vector ${\bf R}_{s}$ was
decayed. Figure \ref{fig:equi}a shows the reorientation in the case of
the 5-1 system which has the longest relaxation time. The second
Legendre polynomial $P_{2}(z)=(3z^{2}-1)/2$ is used for consistency with
analyses further below.   

At this time, approximately, the mean square displacement of a single
monomer begins to coincide with that of the center of mass. For the
longer chains, we first waited until the radius of gyration and the
end-to-end distance did not change systematically any more but
fluctuated only around their mean values. The loss of local
orientation of shorter chain segments is shown in figure
\ref{fig:equi}b. One sees that there are two regimes. On short time
scales ($t^{*}< 5000$), there is a fast decay due to local
processes. On long time scales, however, there is a long tail which is
determined by the overall motion of the whole chain. For the
investigations in the following this overall motion is not important. 

All systems were simulated at least for $t^{*}=20000$. We trust that the
static chain properties were well equilibrated, because the overall
properties like gyration radius settled and at least local orientation
decayed. Moreover, the error estimation for ${\bf R}_{s}$ and
${\bf R}_{g}$ was performed according to a binning analysis
\cite{allen87}. The correlation times for the observed properties
resulting from this analysis were also exceeded substantially. In the
case of 5-1 with 200 monomers, which has the longest equilibration
times, this ``binning time'' is about $t_{b}^{*}=8000$. Such systems
were then simulated for $t^{*}=40000$ to 80000.
\section{Chain Structure in the Melt}
In this section, we investigate the effect of the melt environment
on single chains. The presence of the other chains screens out the 
excluded volume interaction and the chain statistics of a
self-avoiding walk appropriate for chains in good solvent turns into a
simple random walk \cite{Doi:86.1}. This is, for example, evident in the
single chain structure functions (see figure \ref{fig:strfct} in section
\ref{sec:structurefunctions}). In the semiflexible case, one
expects that, at large scales, the Gaussian statistics (random walk) is
fulfilled, whereas on short scales the local stiffness is relevant. Two
concepts can be used for analysis: One is the idea of a Kuhn length
$l_{K}$ which is defined via 
\begin{equation}
  l_{K}=\frac{R_{s}^{2}}{l_{b}(N-1)}.
\end{equation}
This assumes that the melt consists of ``blobs'' of length $l_{K}$ which
contain inside all the local information which is not relevant on the long
scales.  

\begin{figure}
  \begin{center}
    \includegraphics[width=4.5cm,angle=-90]{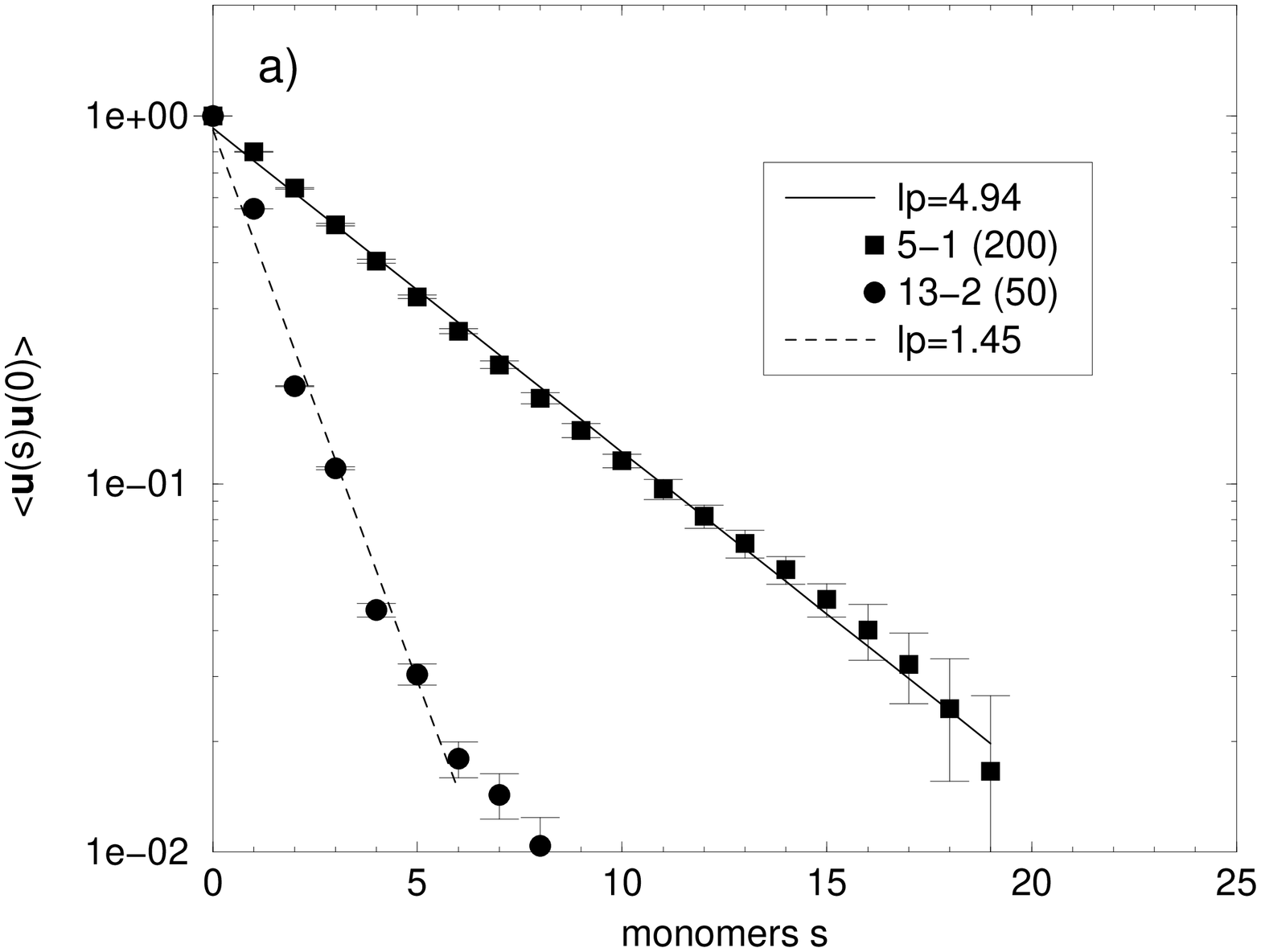}
    \includegraphics[width=4.5cm,angle=-90]{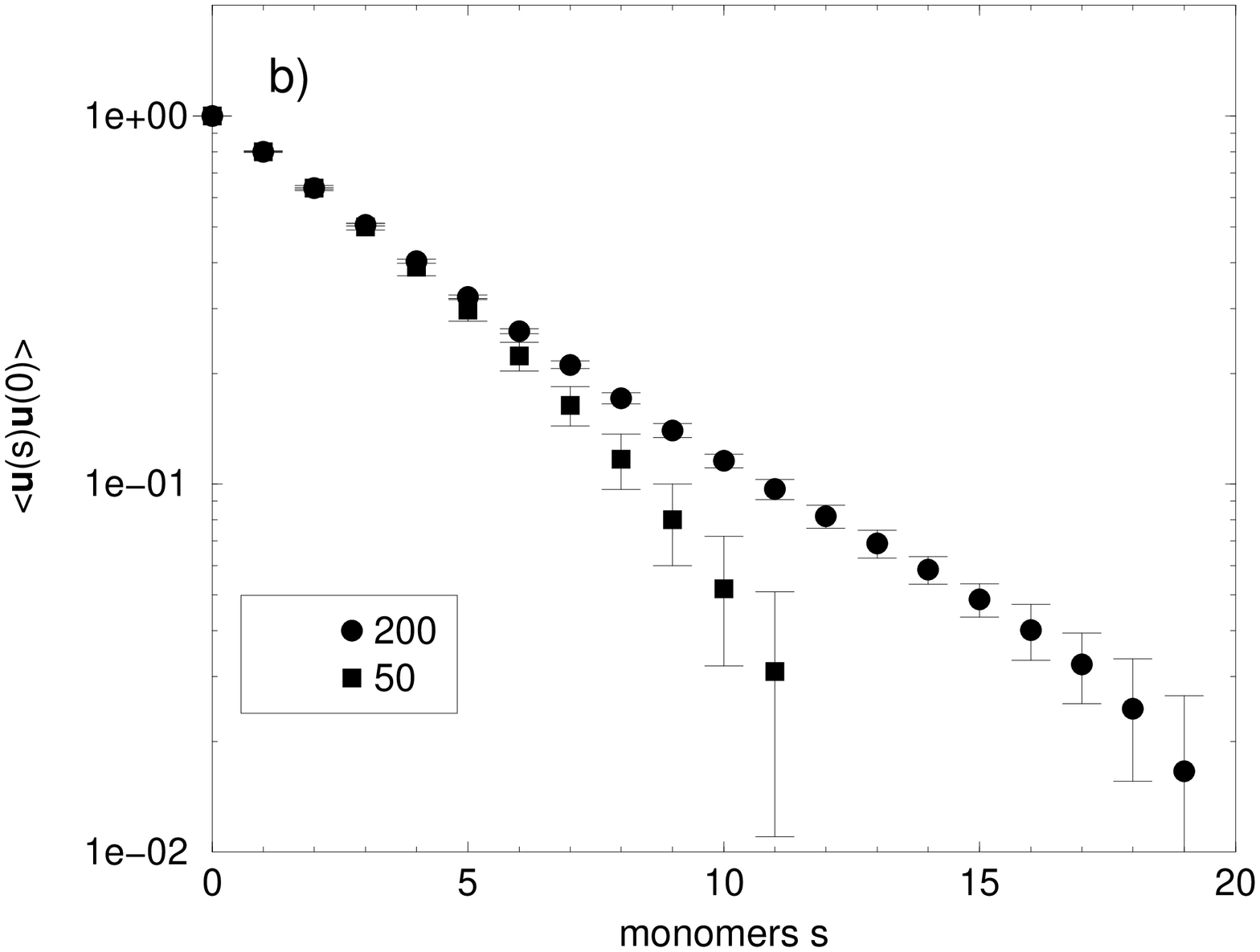}
    \caption{ Bond correlation functions: a) Comparison of systems with uniform
      and alternating stiffness. The lines indicate linear least square fits,
      whose slopes define an effective persistence length. b) Different
      lengths in the 5-1 system. }
    \label{fig:bondcor}
  \end{center}
\end{figure}

The second idea is the persistence length $l_{p}$ which derives from the
worm-like chain model \cite{Doi:86.1,kratky49}. It corresponds to the decay
length of the correlation of bond orientations (the tangent vector)
along the chain.
\begin{equation}
  \langle \cos\alpha(s) \rangle = \langle{\bf u}(s){\bf u}(0)\rangle, \, s:
  \mbox{ monomer index} \label{eq:bondcor}
\end{equation}
which can be shown to decay exponentially in this model
\begin{equation}
  \langle\cos\alpha(s)\rangle=e^{-sl_{b}/l_{p}}.
\end{equation}
Since we do not only apply the bond angle potential to every bond,
but also investigated systems with alternating stiff and flexible
bonds (e.g. 5-2), the persistence lengths of these systems are not a
priori known (at least the $x$-$y$ case $y\ne 1$). In order to
determine $l_{p}$ the bond correlation function (eq. \ref{eq:bondcor}) was
determined, in 100 configurations after the equilibration and the initial
decay was fitted with an exponential $e^{-l/l_{p}}$ (see table
\ref{tab:radius-persist}). If the bending potential was applied to every
monomer the decay was well approximated by an exponential and the decay length
$l_{p}$ was not too far from the expected value $x$ from the bond angle
potential. This is in agreement with Monte Carlo results for stronger
stiffness \cite{kamien97}. In the case of alternating stiffness, minor
deviations from  exponential decay were observed (see figure
\ref{fig:bondcor}a). The error in the bond correlation is about 0.03. Hence,
the systems with very short persistence lengths, were difficult to determine
because only very few points for fitting the decay were available and,
therefore, the resulting error bars are not negligible. However, in all cases
a fit over more than one order of magnitude was possible.  

\begin{table}[htb]
\[
  \begin{tabular}{|c|r|r|r|r|r|}
    \hline
    System & Length &  \multicolumn{1}{c|}{$R_{g}^{2}$} &  
    \multicolumn{1}{c|}{$R_{s}^{2}$} & \multicolumn{1}{c|}{$l_{p}$} &
    \multicolumn{1}{c|}{$l_{K}^{*}$}\\ 
    \hline
    0-1   & 50  & 13.1$\pm$0.2  &  79$\pm$2 & 1.0$\pm$0.1   & 1.68$\pm$0.04\\
    2-1   & 50  & 24.4$\pm$0.2  & 154$\pm$2 & 1.70$\pm$0.01 & 3.27$\pm$0.04\\
    3-1   & 50  & 33.7$\pm$0.2  & 216$\pm$1 & 2.91$\pm$0.05 & 4.59$\pm$0.03\\
    3-1   & 100 & 72.9$\pm$0.2  & 446$\pm$2 & 2.50$\pm$0.01 & 4.70$\pm$0.03\\
    3-2   & 50  & 17.2$\pm$0.2  & 104$\pm$1 & 1.2$\pm$0.1   & 2.21$\pm$0.02\\
    3-2   & 100 & 37.6$\pm$0.1  & 224$\pm$1 & 1.3$\pm$0.1   & 2.35$\pm$0.01\\
    3-2   & 200 & 66.1$\pm$0.1  & 382$\pm$1 & 1.3$\pm$0.1   & 2.00$\pm$0.01\\
    4-2   & 50  & 18.2$\pm$0.1  & 111$\pm$1 & 1.1$\pm$0.1   & 2.36$\pm$0.02\\
    5-1   & 50  &   52$\pm$1    & 357$\pm$3 & 4.01$\pm$0.08 & 7.59$\pm$0.07\\
    5-1   & 100 & 129.9$\pm$0.3 & 833$\pm$3 & 4.71$\pm$0.06 & 8.76$\pm$0.03\\
    5-1   & 200 & 271.8$\pm$0.2 & 1706$\pm$4& 4.94$\pm$0.07 & 8.93$\pm$0.02\\
    5-2   & 50  & 18.8$\pm$0.2  & 114$\pm$2 & 1.2$\pm$0.1   & 2.43$\pm$0.04\\
    5-2   & 100 & 35.3$\pm$0.1  & 203$\pm$1 & 1.2$\pm$0.1   & 2.14$\pm$0.01\\
    5-2   & 200 & 66.9$\pm$0.1  & 394$\pm$5 & 1.35$\pm$0.05 & 2.06$\pm$0.03\\
    13-2  & 50  & 21.4$\pm$0.2  & 133$\pm$2 & 1.45$\pm$0.05 & 2.82$\pm$0.04\\
    13-3  & 50  & 17.0$\pm$0.1  & 103$\pm$1 & 1.0$\pm$0.1   & 2.18$\pm$0.02\\
    100-2 & 50  & 22.7$\pm$0.1  & 142$\pm$1 & 1.51$\pm$0.07 & 3.02$\pm$0.02\\
    \hline
  \end{tabular}
\]
  \caption{Radius of gyration, end-to-end distance (in LJ units), persistence
    length (in monomers) and Kuhn segment length (in LJ units). A $x$-$y$
    system has a stiffening potential of strength $xk_{B}T$ applied every $y$
    monomers. The errors are determined via a binning analysis for error
    estimation \cite{allen87}. }
  \label{tab:radius-persist}
\end{table}

It is not clear if for the $x$-$2$ case the bond correlation has to
follow an exponential law. However, we found this always to be the
case. From a simple argument, the effective persistence length in the
case of persistence lengths $l_{p1}$ and $l_{p2}$ for alternating angles is 
\begin{equation}
  \frac{1}{l_{p}}=\frac{1}{2}\left(\frac{1}{l_{p1}}+\frac{1}{l_{p2}}\right),
\end{equation}
because
\begin{equation}
  e^{-2l/l_{p}}=e^{-l/l_{p1}}e^{-l/l_{p2}}.
\end{equation}
This is exactly true for all points with even monomer distances in the bond
correlation function. This result may be generalized to a repetitive sequence
of $n$ different bond angle potentials
\begin{equation}
  \frac{1}{l_{p}}= \frac{1}{n}\sum_{j=1}^{n}\frac{1}{l_{pj}}.
\end{equation}
A more elaborate calculation in the framework of a generalized
wormlike chain model with varying stiffness yields the same result. 

The persistence length of the fully flexible model is found to be exactly one
monomer distance which is on average $l_{b}=0.97$. The bond correlation
functions (of inner monomers) show in the very beginning a decay with a
persistence length which is close to the expected value (on the length scale
of about 5 monomers). This 
suggests that the very local orientation correlation is determined by the
``true'' potential strength  whereas on longer scales finite size or many
chain effects contribute considerably. This is especially reflected in the
persistence length values for the 5-1 chains (see figure
\ref{fig:bondcor}b), where the bond correlation function for the shorter chain 
shows at distances $s>5$ substantial differences to the longer
chain. They may be attributed to finite chain length effects. All bond 
correlation functions were determined starting from the innermost
monomers in order to avoid end effects as much as possible.

\begin{figure}
  \begin{center}
    \includegraphics[width=5cm,angle=-90]{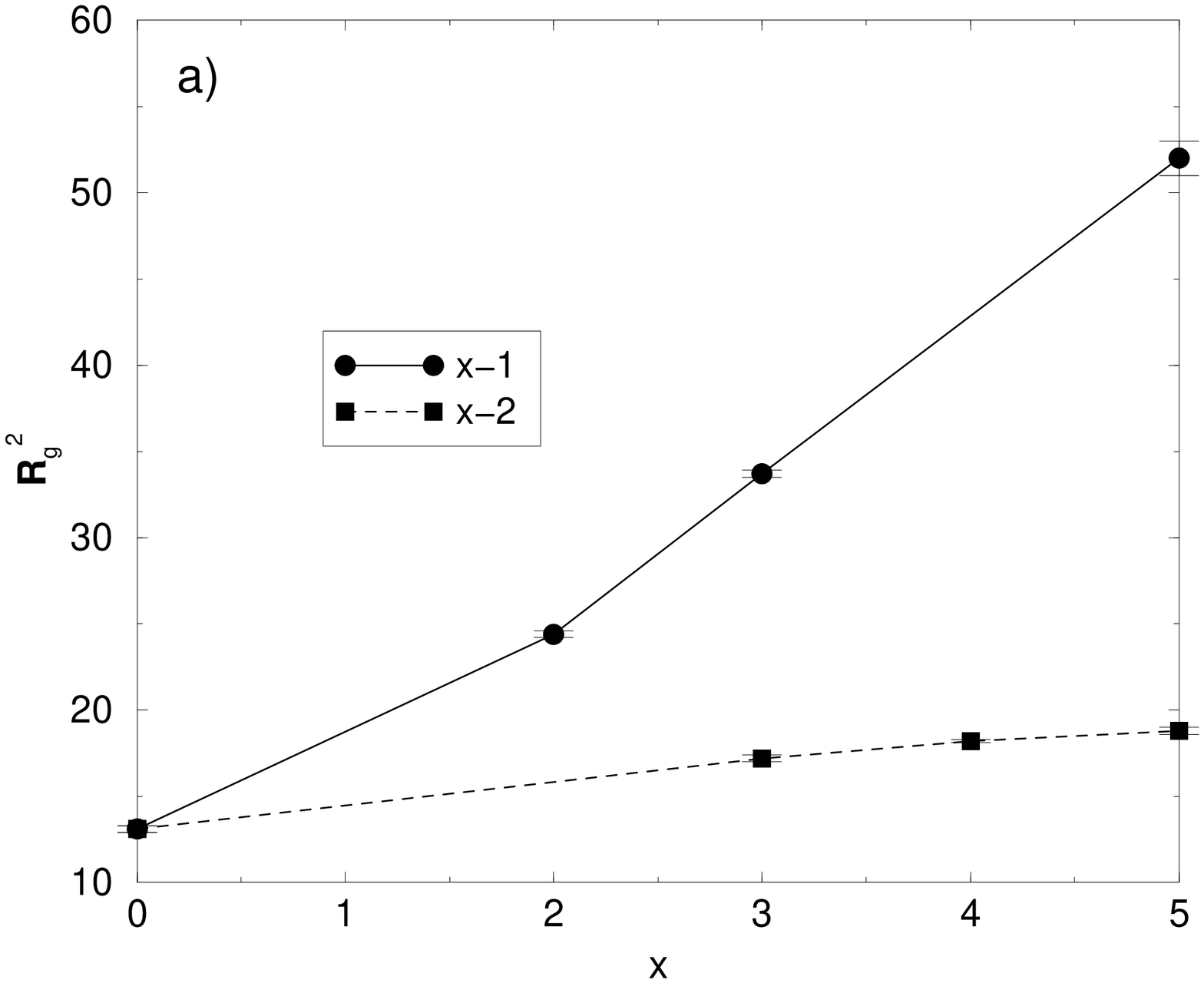}
    \includegraphics[width=5cm,angle=-90]{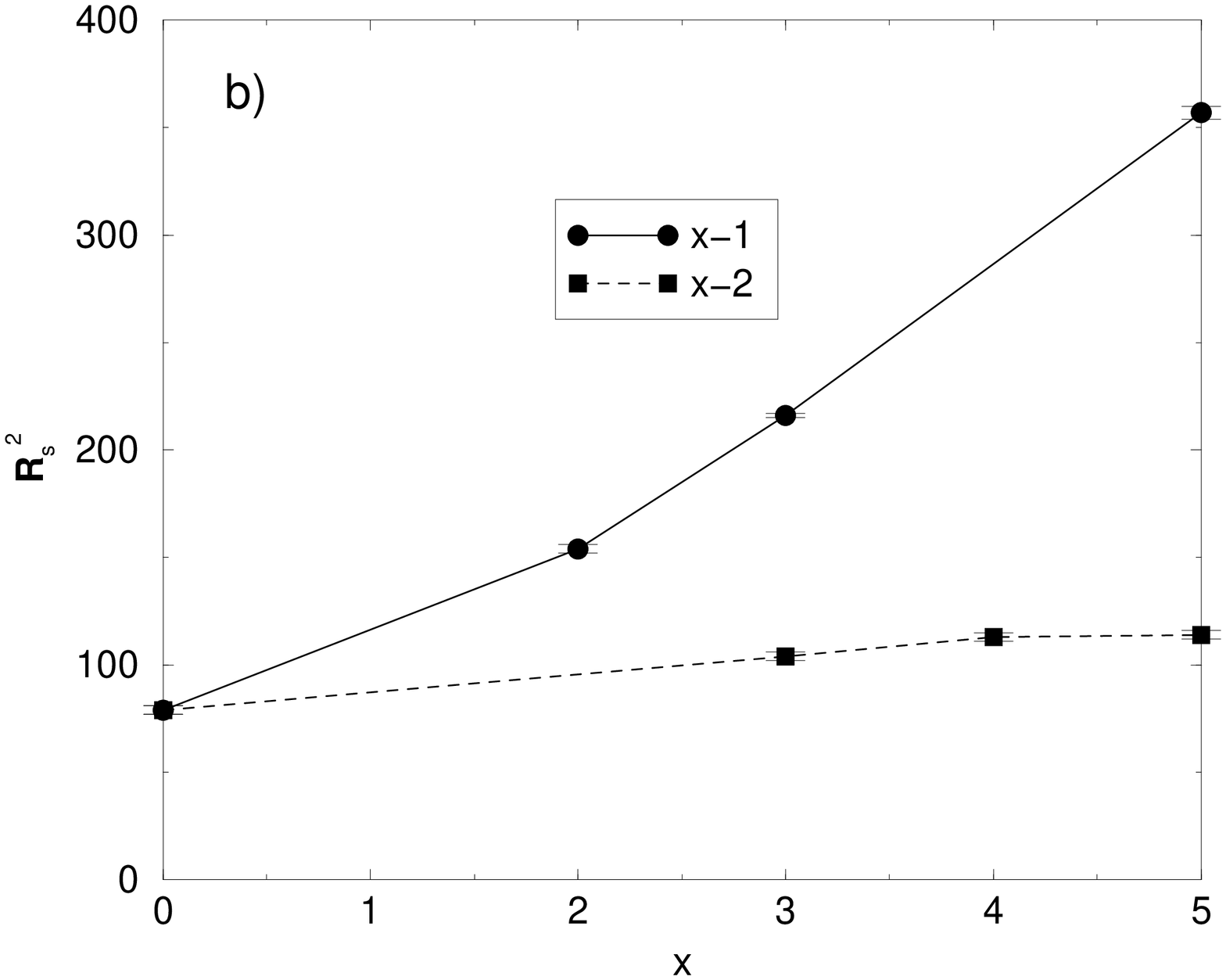}
    \caption{ a) Radius of gyration and b) End-to-end distance for chains of
      length 50 depending on bending strength for $x$-1 and $x$-2 systems
      (uniform and alternating stiffness, resp.). }
    \label{fig:radius}
  \end{center}
\end{figure}

Also the end-to-end distances and radii of gyration of the corresponding
chains were calculated. They are also presented in table
\ref{tab:radius-persist}. Upon increasing $x$, the $x$-1 systems stretch the
chains considerably (see figure \ref{fig:radius}). A much larger bending force
constant is needed for the $x$-2 systems than for $x$-1 if one wants the same
$R_{g}$. Therefore the systems $x$-2 behave more like fully flexible chains
with a bigger monomer which is most strongly seen in the persistence length. 
Note that even in the 100-2 case where an almost rigid and a fully flexible
bond alternate, the chain stretching is not as strong as in the 2-1
case. So there is a fundamental difference between these two scenarios.

At least in the $x$-1 cases we find $2l_{p}\approx l_{K}$ as expected
from the wormlike chain model \cite{Doi:86.1}. The relation
$R_{s}^{2}\approx6R_{g}^{2}$ for the Gaussian chain is well
fulfilled in most of our cases. The larger deviations, e.g. in the 5-1 case
with 50 monomers, may be attributed to finite chain length effects. Therefore,
we do not see substantial deviations from Gaussian behavior. 

\section{Melt structure}
Local orientation of neighboring chains may be measured by the spatial
orientation correlation function OCF. To this end, we define unit
vectors between adjacent monomers
${\bf u}=\frac{{\bf r}_{i}-{\bf r}_{i-1}}{|{\bf r}_{i}-{\bf r}_{i-1}|}$.
The scalar products between two such unit vectors describe the angle
between chain tangent vectors 
\begin{equation}
  \cos\alpha(r)={\bf u}_{chain1}\cdot{\bf u}_{chain2}.
\end{equation}
The distance $r$ denotes the distance between the centers of mass of
the respective chain segments. In order to compare better to NMR
experiments, as well as to avoid the distinction between head and tail
of the chain, we use the second Legendre polynomial
$P_{2}(r)=\frac{1}{2}(3\cos^{2}\alpha(r)-1)$. 

\begin{figure}
  \begin{center}
    \includegraphics[width=4.5cm,angle=-90]{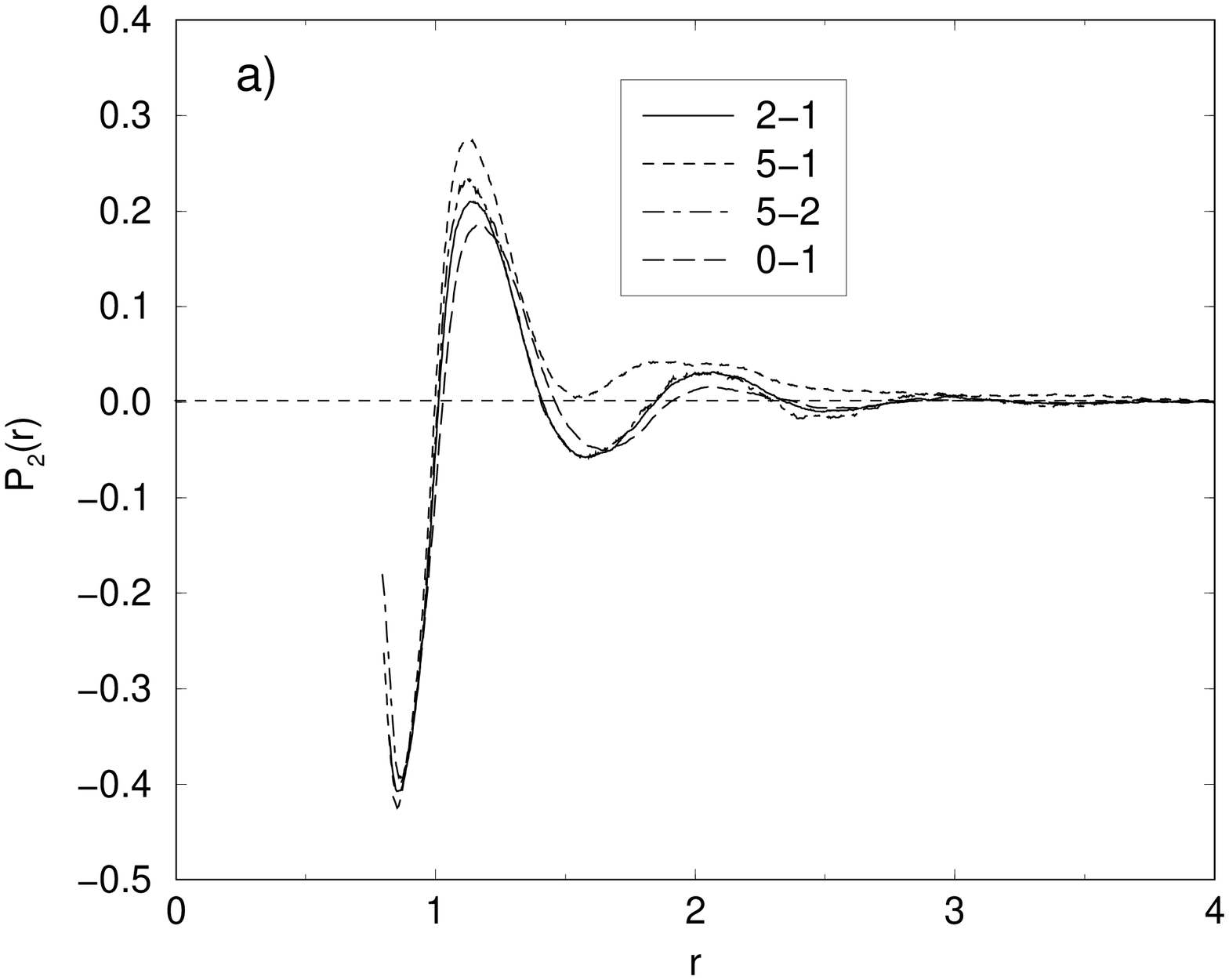}
    \includegraphics[width=4.5cm,angle=-90]{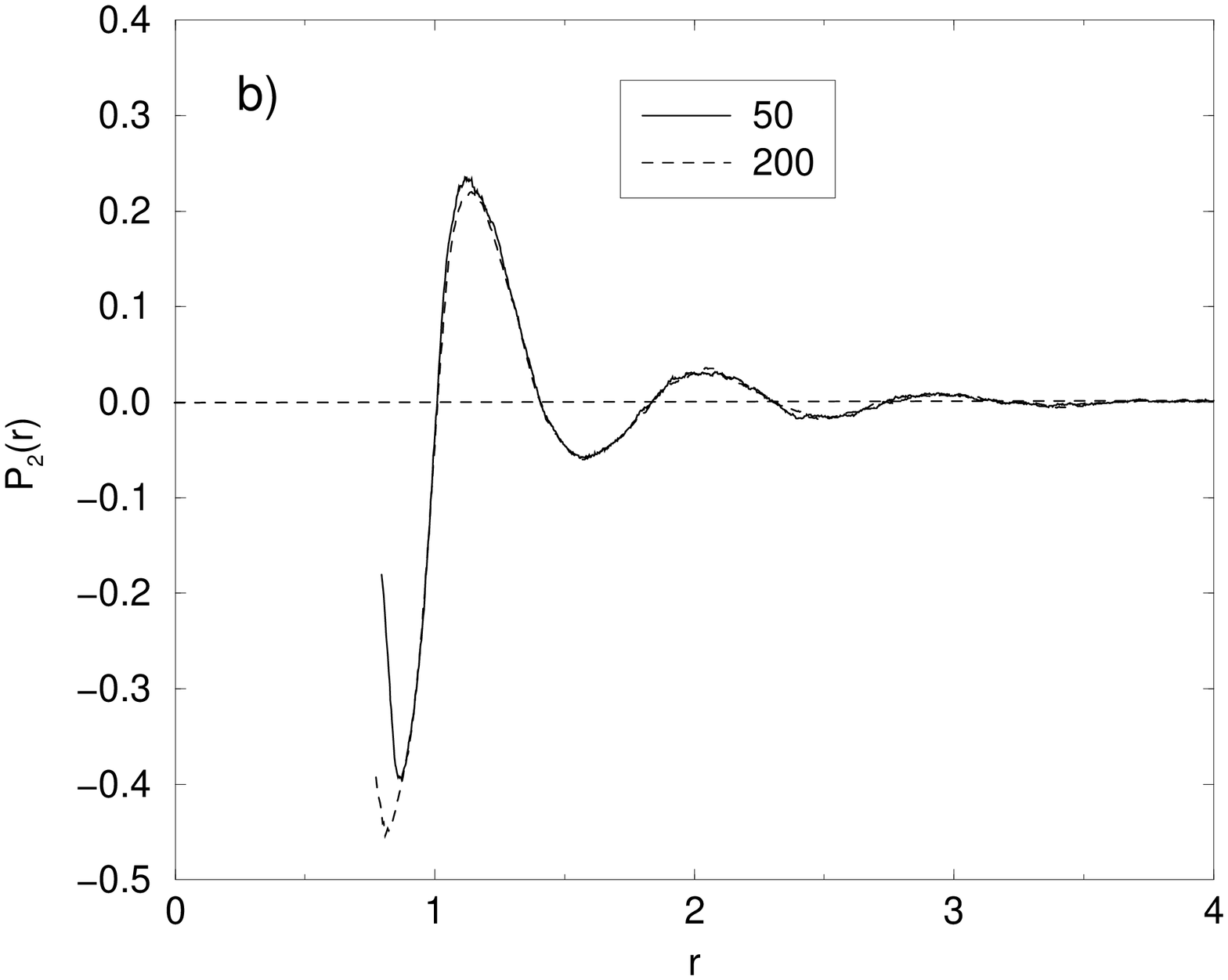}
    \caption{ a) Inter-chain orientation correlation functions for chains with
      50 monomers. b) Inter-chain orientation correlation function for
      different chain lengths for the 5-2 system. In order to better
      distinguish between the lines a running average (over 15 points, $\delta
      x\approx0.05$) was performed.} 
    \label{fig:odf}
  \end{center}
\end{figure}

Figure \ref{fig:odf}a shows inter-chain orientation correlation functions of
different systems. The first minimum $(r<1)$ is close to
$P_{2}=-\frac{1}{2}$ which would indicate a perfect perpendicular
ordering. Two chains which come so close can only pack perpendicular
because of the excluded volume interaction. The radial distribution
function (RDF, see below) shows that there are very few such
contacts. The first peak $(r\approx 1.2)$ shows a preferred
parallel alignment at the distance of the first neighbor. A second parallel
peak follows at $r \approx 2$. The intervening minima $(r\approx
1.6)$ get weaker for stronger orientation which indicates a stronger
local parallel ordering. The OCF decays to zero with $r$ because the
system is globally isotropic, not nematic.

The local ordering is only slightly different for the systems 0-1,
2-1, and 5-2, whereas the 5-1 chain shows a more pronounced local parallel
orientation. For the 5-1 chains, there is residual parallel ordering even at
the intermediate minimum $(r\approx 1.6)$, where the other systems show some
perpendicular ordering. Except for the very few direct contacts, there
is parallel orientation between neighboring chains. This ordering is visible
up to about three monomer diameters. The more flexible systems (0-1, 2-1, 5-2)
show qualitatively a similar ordering, but it is less pronounced and there is
a intermediate preferred perpendicular orientation at the distance of about
the first minimum in the radial distribution function $(r\approx 1.6)$. 

The orientation depends only weakly on the chain length 
(figure \ref{fig:odf}b). This demonstrates that the effect is strictly
local. This holds even though the global dynamics of the chains of
different lengths is strongly different: The chains of length 50 are not yet
entangled whereas the longer chains are already influenced strongly by
entanglements (entanglement length in the 0-1 case $\approx$ 60
monomers \cite{puetzpriv}). 

\begin{figure}
  \begin{center}
    \includegraphics[width=5cm]{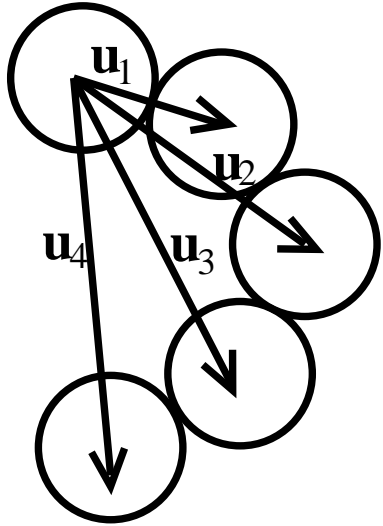}
    \caption{ Definition of unit vectors of bigger segments along the
      chain. Centers of masses are the midpoints of the arrows. }
    \label{fig:expld}
  \end{center}
\end{figure}

Also orientation correlation functions of longer chain segments are
investigated. In this case, not only vectors connecting nearest
neighbors but vectors connecting next-to-nearest neighbor 
beads or beads farther apart are taken into account (see figure
\ref{fig:expld}).  
\begin{equation}
 {\bf u}_{d}:=\frac{{\bf r}_{i}-{\bf r}_{i-d}}{|{\bf r}_{i}-{\bf r}_{i-d}|}
\end{equation}

\begin{figure}
  \begin{center}
    \includegraphics[width=4.5cm,angle=-90]{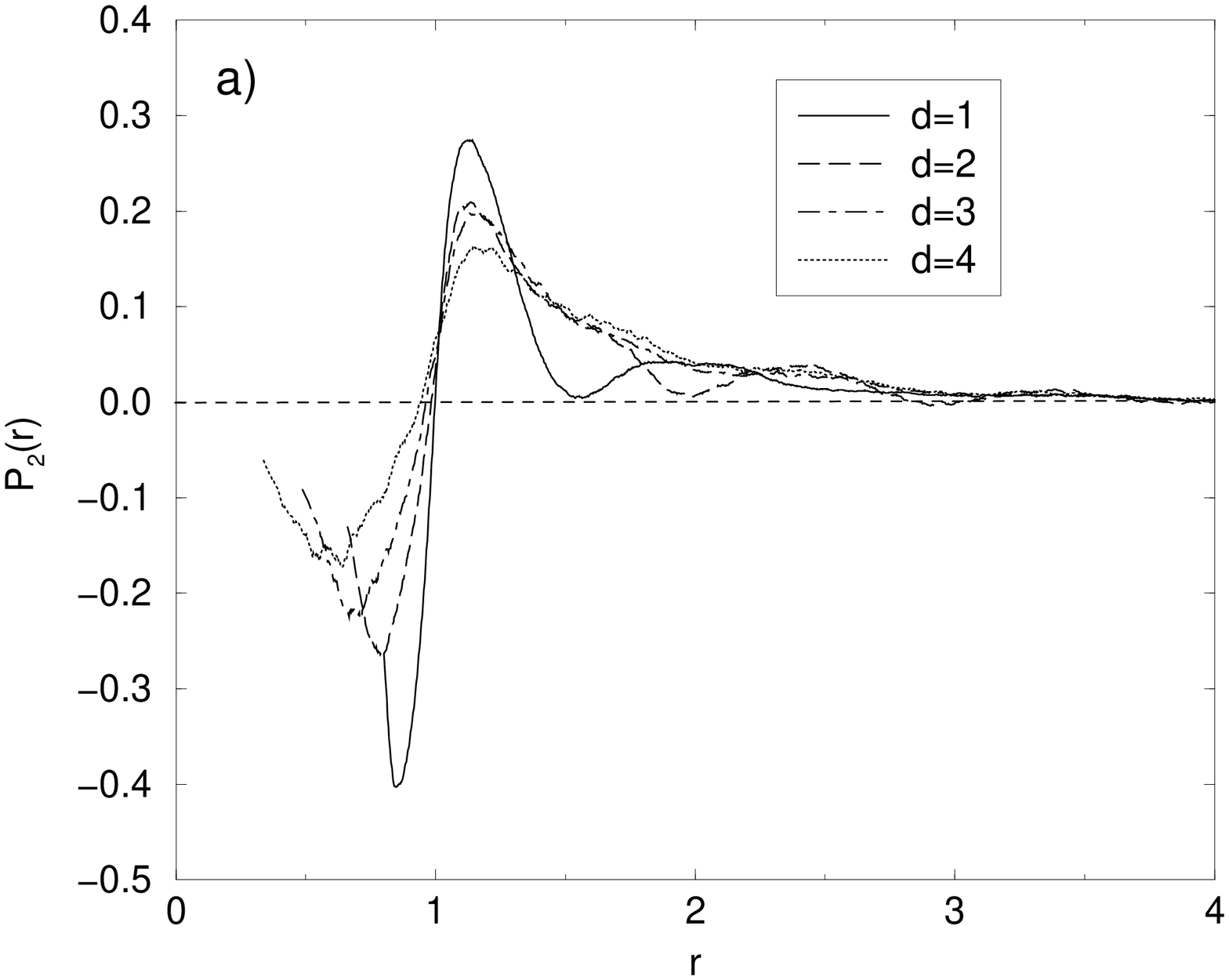}
    \includegraphics[width=4.5cm,angle=-90]{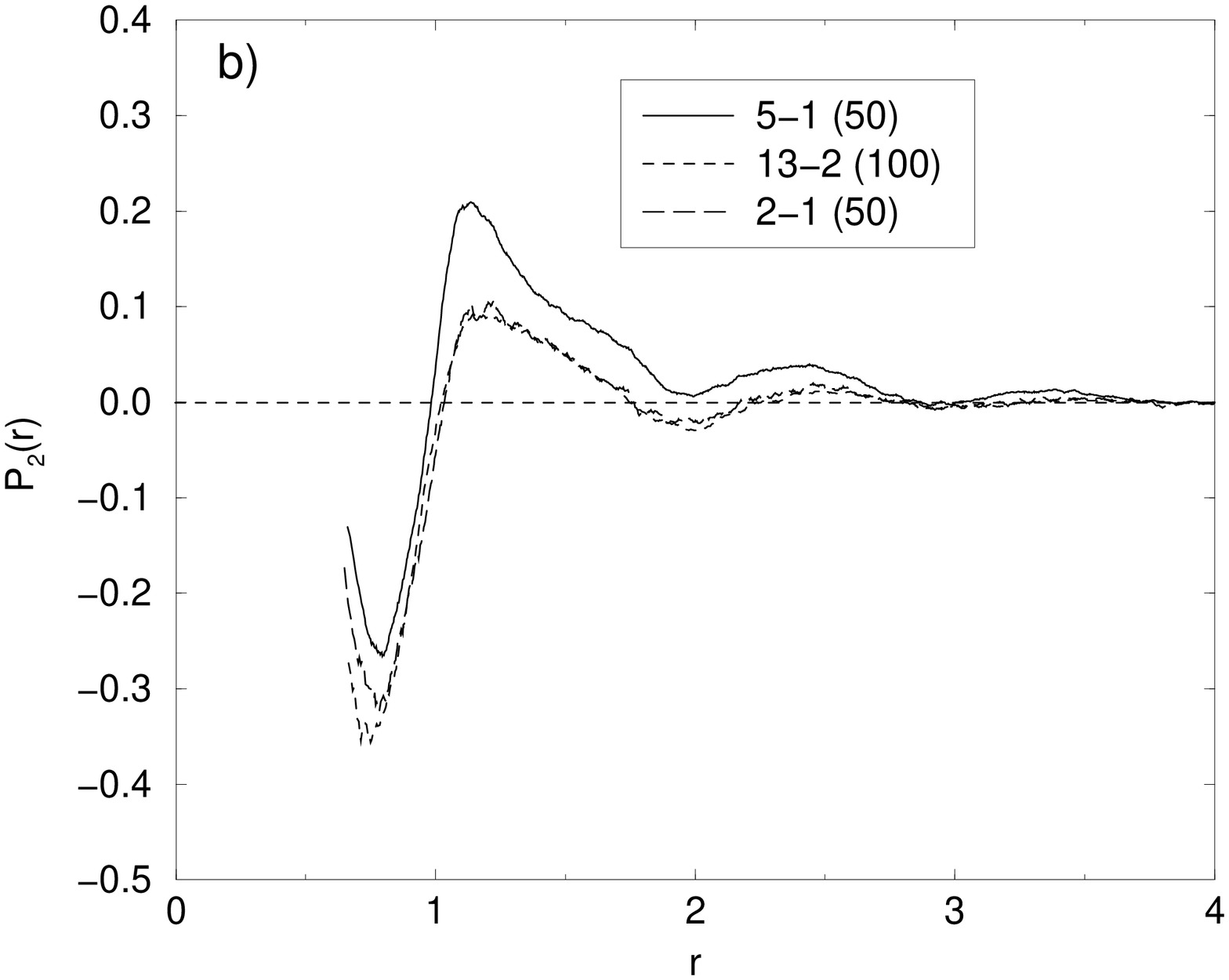}
    \caption{Spatial orientation correlation functions of segments of length
      $d$: a) Different segment lengths in a 5-1 system with 50 monomers. b)
      Different systems, $d=2$. We applied a running average in order to be
      able to see differences of the curves. }
    \label{fig:odfbigseg}
  \end{center}
\end{figure}

It is clear (figure \ref{fig:odfbigseg}a) that the effect of local
parallel chain orientation is not restricted to segments of 2 monomers 
only. It persists when larger chain fragments are analyzed. On the
other hand, the degree of ordering decreases with the segment size considered.
Figure \ref{fig:odfbigseg}b shows again the more pronounced local ordering in
the 5-1 case compared to the more flexible chains. The 2-1 and the
13-2 systems coincide. Their persistence lengths are quite similar, and for
the bigger segment sizes the exact local realization of this persistence
length seems to average out.

Our results are also in qualitative agreement with an early lattice
Monte Carlo investigation of shorter chains \cite{kolinski86}. Lattice
models, however, are biased in favor of orientation correlation.

\begin{figure}
  \begin{center}
    \includegraphics[width=5cm,angle=-90]{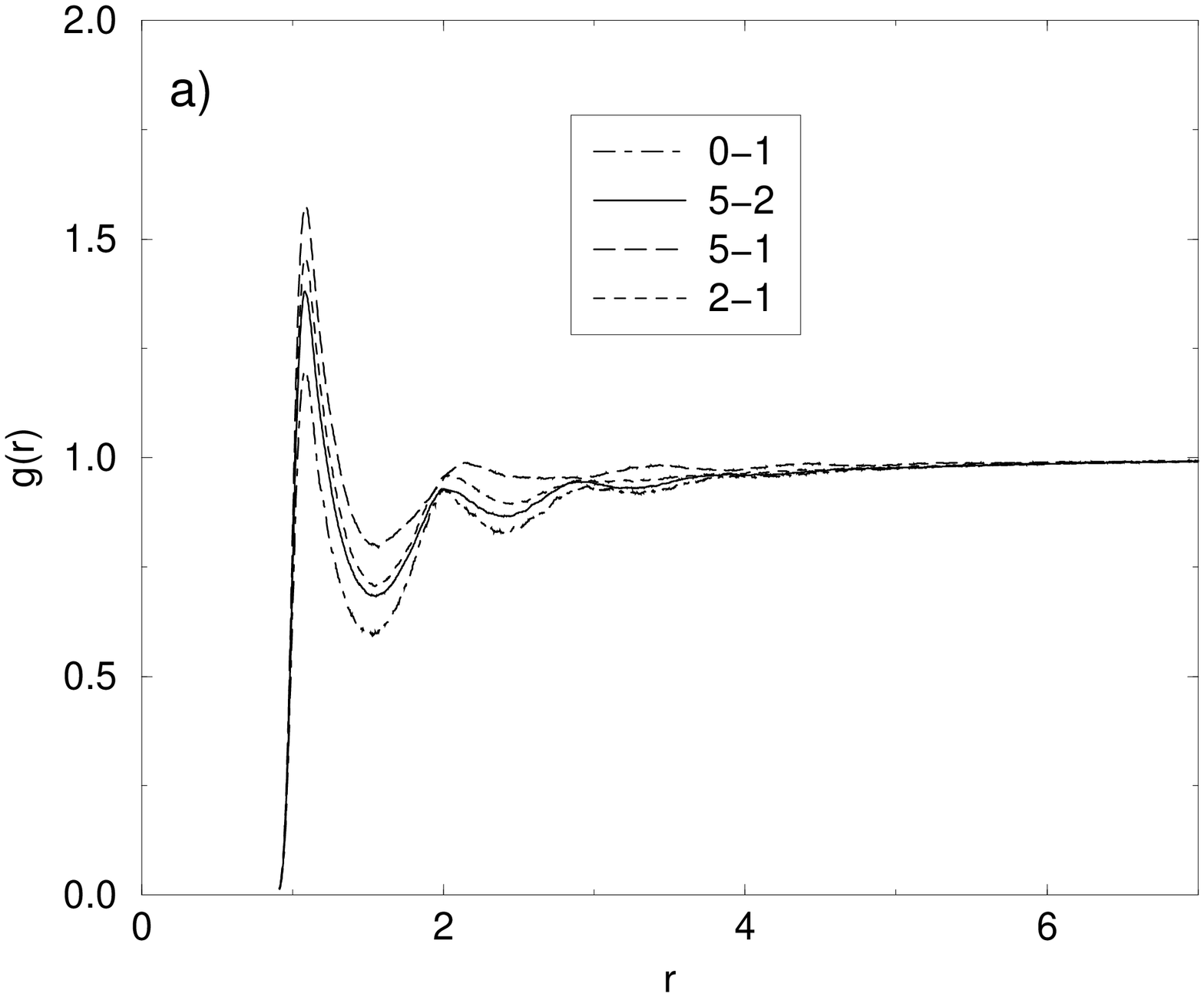}
    \includegraphics[width=5cm,angle=-90]{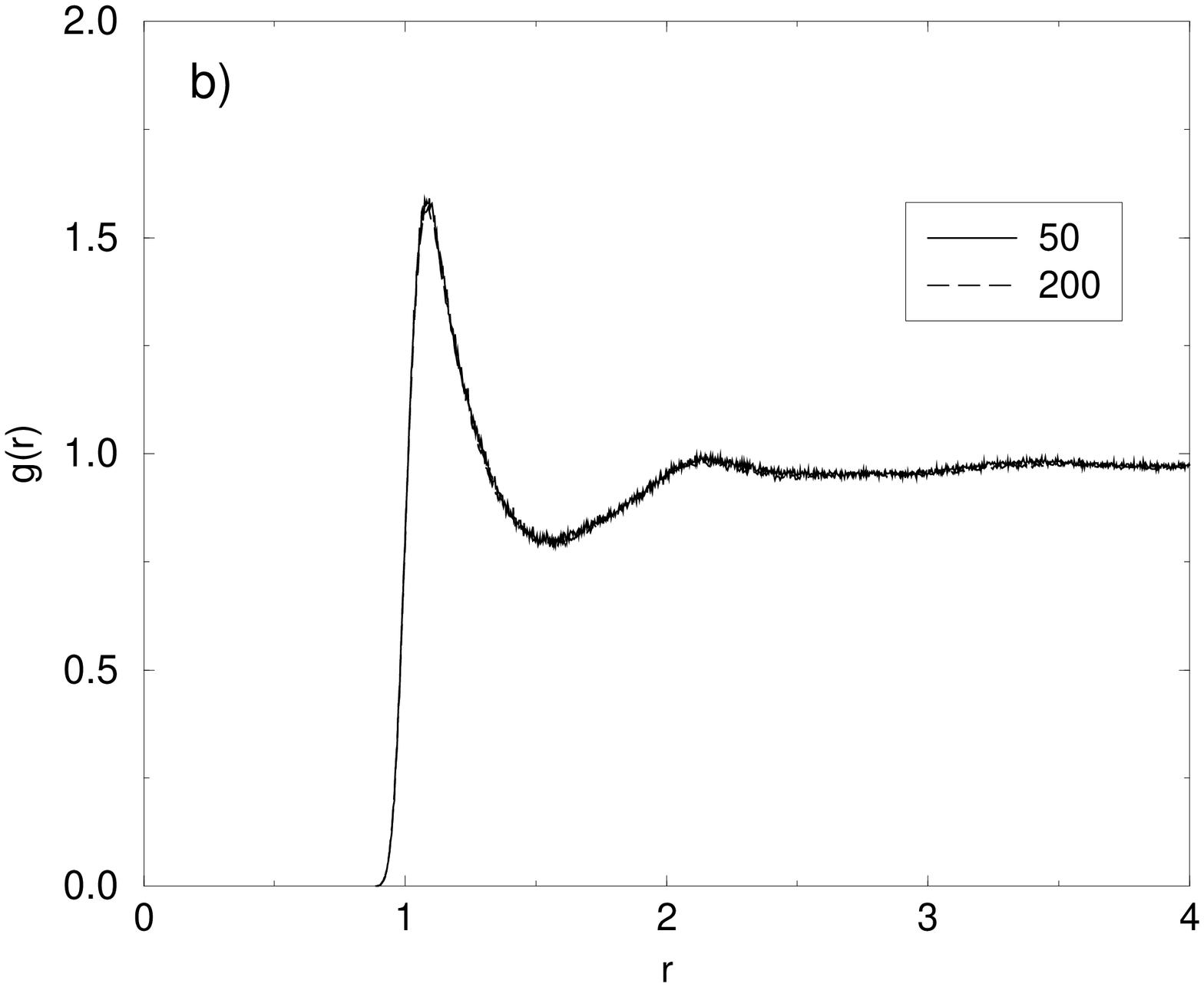}
    \caption{ Inter-chain monomer-monomer radial distribution functions: a)
      Different stiffnesses for length 50. b) Different lengths for 5-1
      system. In order to better distinguish between the curves, a running
      average was applied. }
    \label{fig:rdf}
  \end{center}
\end{figure}

The inter-chain radial distribution function $g(r)$ (RDF) on large
scales does not change much in the case of added stiffness (figure
\ref{fig:rdf}a). However, there are some differences on very local
scales. Both the second and third neighbor peaks are farther apart for
stronger stiffness. Furthermore, the minimum between the first and
second neighbor shell is not as  pronounced as in the more flexible cases.
The local stretching allows a closer approach of chains. This leads to a
reduction of the expected correlation hole. In fully flexible systems
the number of neighbors of one monomer being on the same chain
increase with increasing chain length. No effect of chain length is
seen here (figure \ref{fig:rdf}b) which again reflects the strict
locality of the structure formation in the melt. 
\section{Structure functions}
\label{sec:structurefunctions}

\begin{figure}
  \begin{center}
    \includegraphics[width=4.5cm,angle=-90]{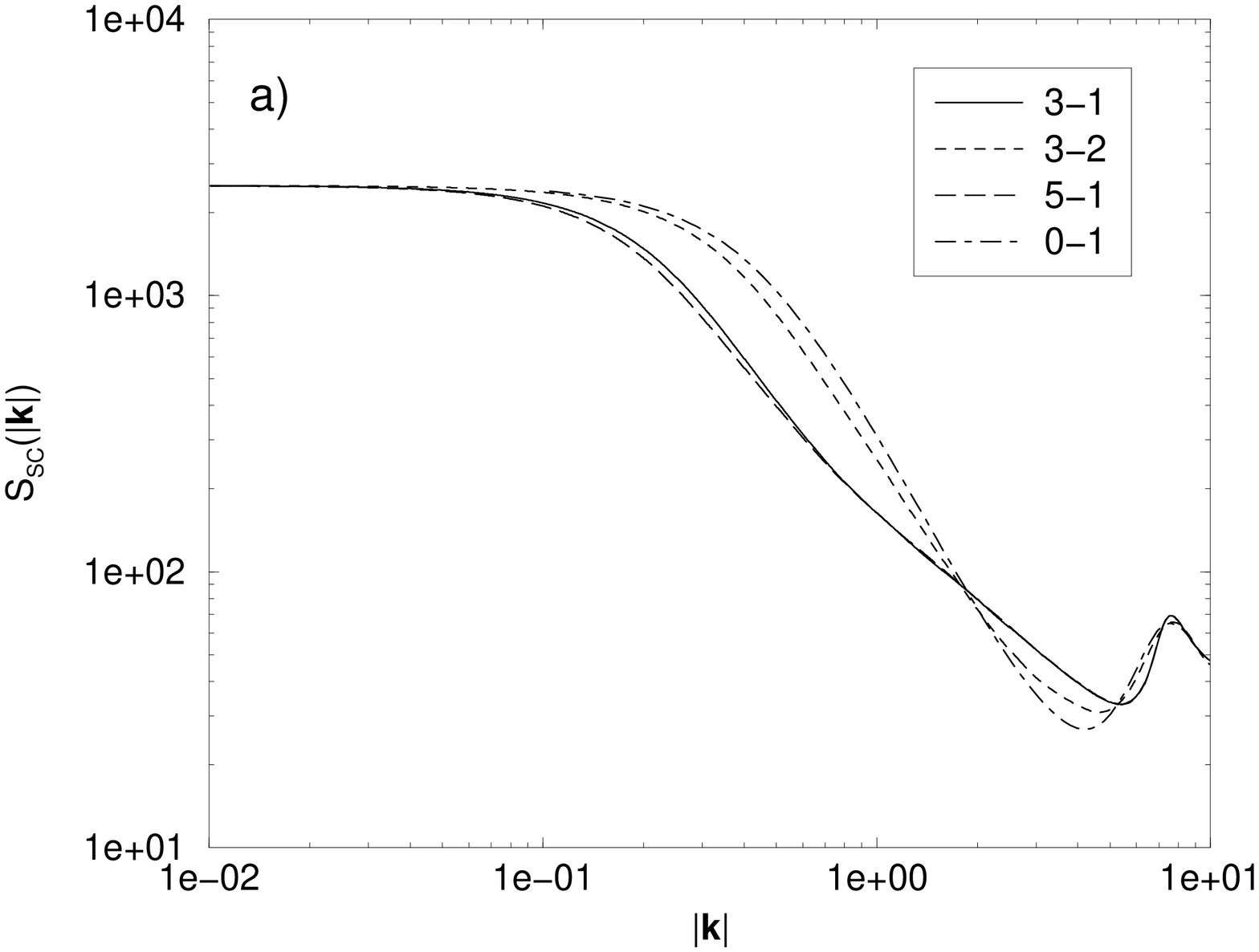}
    \includegraphics[width=4.5cm,angle=-90]{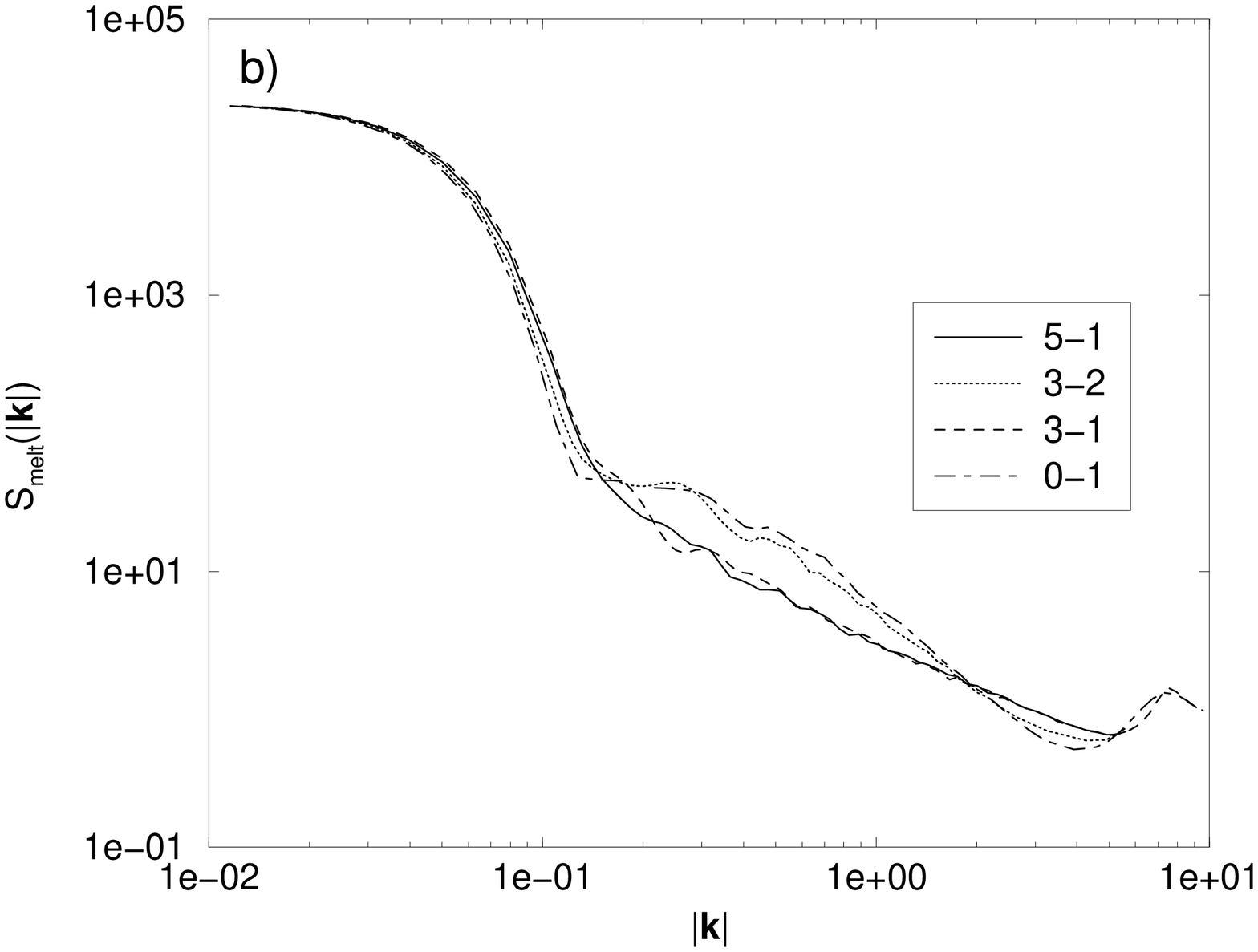}
    \caption{ Structure functions for chains of 50 monomers: a) Single-chain
      structure function. b) Melt structure function.}
    \label{fig:strfct}
  \end{center}
\end{figure}

The structure of single chains and of the overall melt may be additionally
characterized by static structure functions. Figure \ref{fig:strfct} shows the
single-chain and melt structure functions of our systems. The (isotropically
averaged) melt structure function is defined as  
\begin{equation}
  S_{melt}(k) = \frac{1}{N}\langle|\sum_{m=1}^{N_{C}}\sum_{j=1}^{n_{b}}
  \exp(ikr_{j}^{m})|^{2}\rangle = S_{SC}(k)S_{inter}(k)
\end{equation}
where $S_{SC}$ denotes the single chain structure function
\begin{equation}
  S_{SC}(k) = \frac{1}{N}\sum_{m=1}^{N_{C}}\langle|\sum_{j=1}^{n_{b}}
  \exp(ikr_{j}^{m})|^{2}\rangle.
\end{equation}
The first sums run over all chains ($N_{C}$: number of chains, $m$:
chain index), the second along the chains ($n_{b}=N/N_{C}$: number of
beads along the chain, $j$: monomer index along the chain). 

In the limit for $k\to 0$ the chain structure is no more visible but
we just see a massive object which is related to the first plateau in
$S_{SC}$. The next ``scaling'' regime is connected to the fractal
nature of the chains. The self similarity yields a decay with
$k^{-\frac{1}{\nu}}$, where $d=\frac{1}{\nu}$ is the fractal dimension
of the chain. For a Gaussian chain we have $\nu=\frac{1}{2}$. Stretching leads
to a smaller fractal dimension resulting in a less steep decay. In the
large-$k$ range we deal with structure of the size of one monomer. A bead
spring model has no structure on a shorter scale. The melt structure function
is the Fourier transform of the density-density correlation function. It shows
therefore some additional peaks which correspond to peaks in the RDF. Hence,
it contains not only information about the single-chain structure but also
about the overall structure of the whole system.

The single-chain structure functions $S_{SC}$ of the $x$-1 systems look very
similar, whereas there are strong differences between the 3-1 and the 3-2
system, the latter behaving very much like a fully flexible
system. The crossover to the scaling regime is shifted to smaller
$k$-vectors for stiffer chains. There is additionally a crossover
($k=0.8$) between two regimes in the decay which means that there is
different fractal chain structure on different length scales. At larger scales
(small-$k$ regime) the slope does not differ much from the fully flexible
case, whereas on intermediate scales larger deviations occur, which indicate
local chain stretching. The 3-2 system, however, is close to the fully
flexible (Gaussian) system. But its slope is not as steep, which hints at a
slight stretching of the chains compared to the Gaussian chain. This minor
difference between the alternating stiffness and the fully flexible chains
supports our earlier suspicion that the alternating chains behave like
renormalized bead-spring chains with larger effective monomers. On the
other hand, the stiffened chains with true semiflexibility are
strongly different on intermediate scales.

The melt structure functions $S_{melt}(k)$ differ also between the $x$-1
case and the $x$-2 case. The latter is again very similar to the 0-1
system. The slopes in the regimes around $k\approx 1$ are clearly
different, whereas the fine structure revealing peaks connected to the
neighboring shells is quite similar. The exact positions of these peaks
differ, however, which shows again that the distance to the nearest neighbors
is slightly altered with stiffness. Figure \ref{fig:strfct}b shows
that all differences are on local scales $(k>0.1)$. The static structure
functions coincide for $k\to 0$, where the overall structure on large scales
gets important. So the melt structure function shows that all systems
behave similar on large scales  but the systems with homogeneous and
alternating stiffness differ on local scales.  
\section{Conclusions}
The static structure of semiflexible polymers in the melt was
determined. The stiffness strongly affects the persistence length,
end-to-end distance and radius of gyration. Stiff chains are more
stretched than chains only interacting via excluded volume. This also
affects the local mutual ordering of the chains. Stiff chains pack
more parallel on local scales whereas the overall structure remains
isotropic. The chain length does not influence this strictly local
phenomenon. Systems consisting of alternating stiff and flexible links
behave similar to systems with much weaker overall persistence
lengths. Their structure is similar to the structure of fully
flexible chains with larger monomers. Their overall persistence length is
smaller than expected by a analytical calculation using the persistence
lengths of the respective potentials. Finally, the overall local structure of
the melt differs considerably for alternating and homogeneous stiffness. 

The static data presented here shows already ordering effects but its effect
on dynamical properties relevant to NMR experiments can not be inferred. In
order to compare directly to NMR experiments on real polymer melts, dynamical
investigations are needed. Such simulations and analyses are presently being
performed with the mesoscopic model of this article. Moreover, detailed
atomistic simulations are underway for melts of specific polymers. They
provide directly the time evolution of the atom-atom vectors monitored in the
experiments, at least for short times. These simulations will be mapped onto
more coarse grained simulations like the one presented here. This mapping will 
allow us to make the connection between the $(x,y)$ parameters of our model
and real polymers. The presented static properties are a first step on the way
of understanding ordering phenomena as examined in NMR experiments. 
\section*{Acknowledgements}
Valuable discussions with Andreas Heuer, Mathias P\"utz, and Heiko Schmitz are
gratefully acknowledged.  

\end{document}